\documentclass[aip,reprint,amsmath,amssymb]{revtex4-1}

\usepackage{graphicx}
\usepackage{dcolumn}
\usepackage{bm}

\usepackage[utf8]{inputenc}
\usepackage[T1]{fontenc}
\usepackage{mathptmx}
\usepackage{etoolbox}

\makeatletter
\def\@email#1#2{%
 \endgroup
 \patchcmd{\titleblock@produce}
  {\frontmatter@RRAPformat}
  {\frontmatter@RRAPformat{\produce@RRAP{*#1\href{mailto:#2}{#2}}}\frontmatter@RRAPformat}
  {}{}
}%
\makeatother
\begin{document}

\title{Magnetoreception in cryptochrome enabled by \\one-dimensional radical motion}
\author{Jessica L.\ Ramsay}
\author{Daniel R.\ Kattnig$^{*,}$}
\email{d.r.kattnig@exeter.ac.uk}
\affiliation{%
 Living Systems Institute and Department of Physics, University of Exeter, Stocker Road, Exeter, Devon, EX4 4QD, United Kingdom
}%

\date{\today}

\begin{abstract}
    A popular hypothesis ascribes magnetoreception to a magnetosensitive recombination reaction of a pair of radicals in the protein cryptochrome. Many theoretical studies of this model have ignored inter-radical interactions, particularly the electron-electron dipolar coupling (EED), which have a detrimental effect on the magnetosensitivity. Here, we set out to elucidate if a radical pair allowed to undergo internal motion can yield enhanced magneto-sensitivity. Our model considers the effects of diffusive motion of one radical partner along a one-dimensional reaction coordinate. Such dynamics could in principle be realized either via actual diffusion of a mobile radical through a protein channel, or via bound radical pairs subjected to protein structural rearrangements and fluctuations. We demonstrate that the suppressive effect of the EED interactions can be alleviated in these scenarios as a result of the quantum Zeno effect and intermittent reduction of the EED coupling during the radical's diffusive excursions. Our results highlight the importance of the dynamic environment entwined with the radical pair and ensuing magnetosensitivity under strong EED coupling, where it had not previously been anticipated, and demonstrate that a triplet-born radical pair can develop superior sensitivity over a singlet-born one.
\end{abstract}

\maketitle

\section{Introduction}

\begin{figure}[tbhp]
	\centering
	\includegraphics[width=7.8cm]{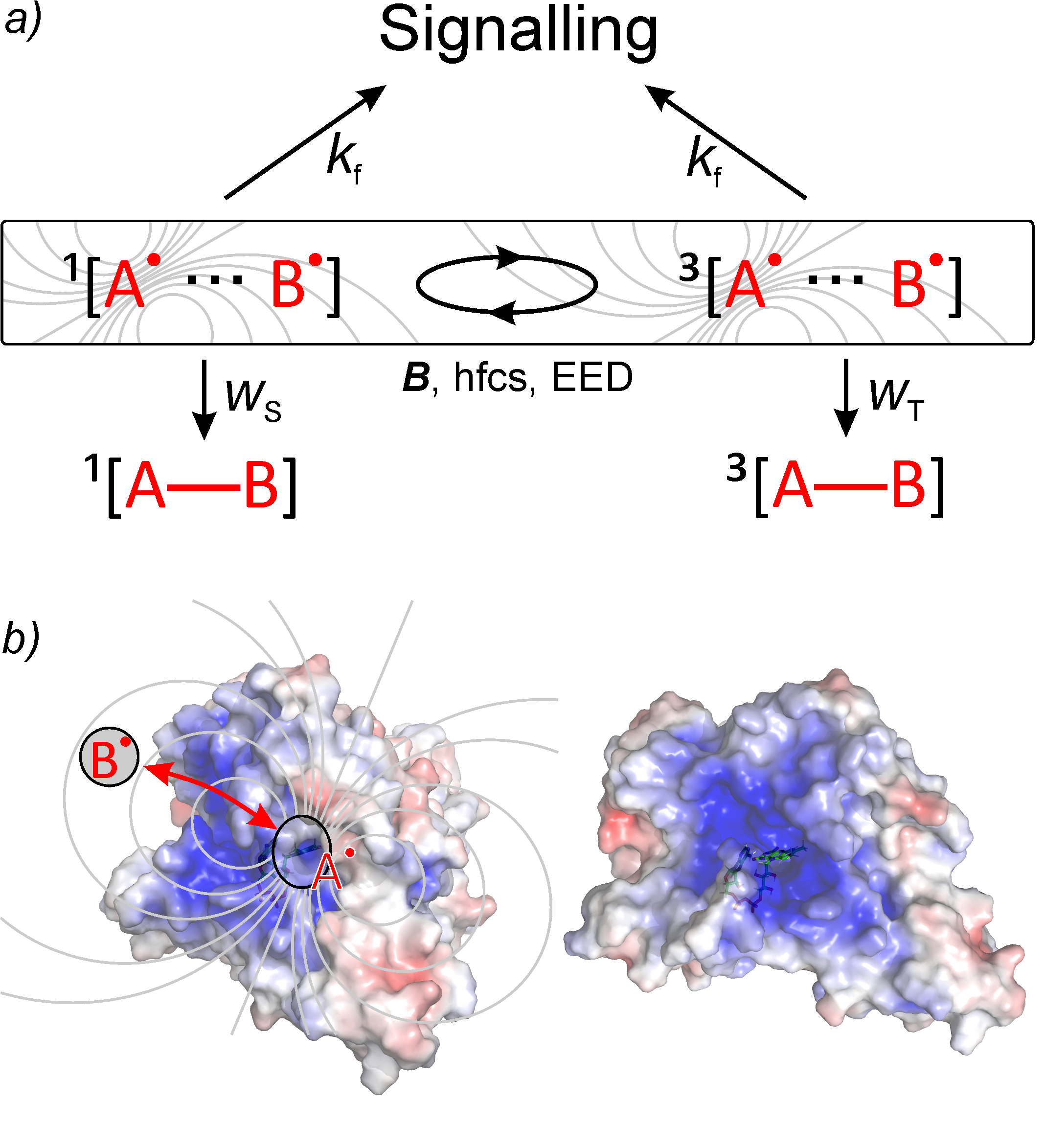}
	\caption{a) Radical pair reaction scheme leading to magnetosensitive yields of the signalling state. The radical pair comprising radicals A$^{\bullet}$ and B$^{\bullet}$ coherently interconverts in the basis of singlet and triplet electronic states as a result of hyperfine interactions (hfcs), the Zeeman interaction with the applied magnetic field $\mathbf{B}$, and the electron-electron dipolar coupling (EED; field lines schematically indicated for A$^{\bullet}$ by gray lines). The superscript numbers indicate the electronic spin multiplicity. The radical pair can spin-selectively react to form products from the singlet (distant dependent rate $w_S$) and triplet (rate $w_T$) manifold, the reaction products being denoted as $^1[\mathrm{A}-\mathrm{B}]$ and $^3[\mathrm{A}-\mathrm{B}]$, respectively. In addition, the system can give way to signalling to downstream processes (rate constant $k_f$), which is thought to be induced by spin-independent structural rearrangements of the protein \cite{schuhmann2021exploring}). b) Representation of the pigeon cryptochrome 4 (based on PDB-ID 6PU0; representative structure from MD simulation with phosphate-binding loop reconstructed taken from \cite{schuhmann2021exploring}). The protein structures were visualized in the PyMOL Molecular Graphics System, Version 2.5.0 Schrödinger, LLC \cite{PyMOL}. The surface shows the electrostatic potential of the protein, with blue highlighting areas of positive potential. A mobile radical B$^{\bullet}$ can approach the FAD co-factor, supposed to form radical A$^{\bullet}$, from the indicated site via a pore like structure. \emph{Left}: side view; \emph{right}: front view. The non-covalently bound FAD co-factor is shown inside of the protein.}
	\label{fig:reaction_scheme}
\end{figure}

Magnetoreception, the ability to sense Earth's magnetic field, is widespread across the animal kingdom \cite{wiltschko1995migratory, mouritsen2018long}. While the phenomenon has been well documented, frustratingly little is known about the mechanistic underpinnings, the sensor structures and downstream reaction pathways \cite{johnsen2008magnetoreception,nordmann2017magnetoreception,mouritsen2018long}. 
A light-dependent inclination compass has often been documented and is particularly well researched for migratory songbirds \cite{wiltschko1972magnetic,hore2016radical}. Magnetosensitive traits have also been widely studied in fruit flies, revealing some mechanistic details and informing possible transduction pathways \cite{bradlaugh2021exploiting,gegear2008cryptochrome,fedele2014genetic}. The most promising hypothesis to explain these forms of magnetoreception is based on the radical pair mechanism (RPM), as suggested by Schulten \emph{et al.}\ in 1978 and later refined by Ritz \emph{et al.}\ \cite{schulten1978biomagnetic,ritz2000model,wiltschko1995migratory,mouritsen2018long}.
The hypothesis assumes a magnetically sensitive chemical reaction in the protein cryptochrome. There, a light-dependent electron transfer reaction is thought to generate a spin-correlated radical pair, which acquires magnetosensitivity as a result of the interplay of coherent quantum spin dynamics and the spin-selective chemical reactivity of the radical pair. As a consequence, the yield of a signalling state is predicted to depend on the intensity and direction of the applied magnetic field.  A promising candidate for such a magnetosensitive radical pair is the flavin adenine dinucleotide/tryptophan pair (FAD$^{\bullet-}/$W$^{\bullet+}$) in cryptochrome, which is produced by a sequence of electron transfers along three or four tryptophan residues of a conserved electron transfer chain \cite{xu2021magnetic, hochstoeger2020biophysical}. Some evidence also points to magnetosensitivity in the dark-state re-oxidation of the fully reduced FAD accumulated via an upstream, light-dependant reaction route \cite{wiltschko2016light, pooam2019magnetic}. This phenomenon has been implicated with the flavin semiquinone/superoxide radical pair (FADH$^{\bullet}/$O$_{2}^{\bullet-}$), but questions remain as to the fast spin relaxation of the latter \cite{ritz2009magnetic,solov2009magnetoreception,hogben2009possible,wiltschko2016light,muller2011light,player2019viability, mondal2019theoretical}.  In addition, tyrosyl (Y$^\bullet$) and ascorbyl radicals (A$^{\bullet-}$) have been considered in theoretical studies \cite{lee2014alternative, atkins2019optimal, deviers2022anisotropic}. In fruit flies, the 52 C-terminal amino acids of cryptochrome appear sufficient to mediate magnetic field sensitivity, suggesting a non-canonical, potentially protein-associated radical pairs, likely involving FAD and a yet unknown electron donor \cite{bradlaugh2021exploiting, bradlaugh2021essential,fedele2014genetic}. Figure \ref{fig:reaction_scheme} provides a simplified reaction scheme, as it applies to this study; the FAD-radical is denoted as A$^{\bullet}$ here. Figure S1 in the Supporting Information illustrates the cryptochrome redox-cycle comprising photo-reduction and re-oxidation, from which the abovementioned radical pairs are assumed to derive. Radical pair magnetoreception has been the subject of many review articles, which we refer the reader to for additional insights and details \cite{wiltschko2019magnetoreception,kim2021quantum,hore2016radical, bradlaugh2021exploiting}.

The principle possibility of magnetosensitivity of a radical pair reaction in the geomagnetic field has been established in a multitude of theoretical studies, but compelling experimental evidence is still lacking for cryptochrome in the geomagnetic field. In any case, the process requires that certain conditions must be met \cite{hore2016radical}: a) the radical pair lifetime must be sufficient (of the order of at least one Larmor precession period in the geomagnetic field, i.e.\ $\approx 1\mskip3mu\mu$s); b) the reaction dynamics must involve competition between at least two reaction channels in a spin-dependent fashion; c) spin relaxation times ought to permit at least one Larmor precession and should ideally exceed the radical pair lifetime; and d) the radicals must not be too close or too far apart. The latter requirement reflects the necessity that the radicals undergoes a spin-selective recombination reaction, whilst ideally be spared from inter-radical coupling, such as the exchange and electron dipolar (EED) coupling. The overarching consensus so far is that the simple RPM based model is able to provide a tentative explanation of many of the traits of magnetoreception. Yet, conceptual issues remain. For realistic systems, the above mentioned prerequisites are challenged due to environmental noise that limits the coherent lifetime \cite{kattnig2016electron} and due to the inter-radical coupling at reasonable recombination distances ($\approx 1.5\mskip3mu$nm) \cite{babcock2020electron, efimova2008role}. While the exchange interaction is thought to be negligible for prototypical radical pairs in cryptochrome \cite{nohr2017determination}, the EED coupling distinctly manifests, as it decays slowly with the inter-radical distance $r$, i.e.\ as $r^{-3}$ (as opposed to the exchange interaction, which is characterised by an approximately exponential distance dependence) \cite{efimova2008role}. At a typical distance of $r = 1.5\mskip3mu$nm, the dipolar coupling for example amounts to $d = - \frac{\mu_{0} \gamma_{e}^2 \hbar}{8\pi^2r^{3}} \approx -15\mskip3mu$MHz, which exceeds the Larmor precession frequency ($1.4\mskip3mu$MHz in a $50\mskip3mu\mu$T magnetic field) by an order of magnitude. Significant inter-radical interactions are however detrimental to the magnetosensitivity in weak magnetic fields, as the interactions tend to lock the system in the singlet and triplet manifolds, suppress their coherent interconversion and thus the magnetic sensitivity \cite{babcock2020electron,efimova2008role,dellis2012quantum}. 

The effect of inter-radical interactions on the magnetosensitivity of cryptochromes in weak magnetic fields has only gained momentum recently \cite{babcock2020electron,keens2018magnetosensitivity}. The majority of the theoretical studies so far have neglected the inter-radical couplings. The mutual compensation of the exchange and EED interactions ($J/D$ compensation) was suggested as a rational to do so \cite{efimova2008role}, but although functional in principle, this effect was not found effective in radical pairs involving the flavin radical \cite{babcock2020electron}. Three-radical models have been able to provide larger magnetosensitivity in the presence of EED coupling by suitable placement of an inert radical bystander \cite{babcock2020electron} or by postulating a spin-selective recombination reaction involving a radical of the primary pair and a ``scavenger'' radical \cite{babcock2021radical}. While successful in principle, the resulting reaction schemes and biological implementation are complex and currently not supported by direct evidence (but direct evidence is scare also in relation to the simpler models). Kominis and co-worker have pointed out that the quantum Zeno effect, i.e.\ strongly asymmetric recombination, can in principle immunizes a radical-pair sensor to the deleterious inter-radical interaction, which they have demonstrated for the simple model of an exchange-coupled one-proton radical pair subject to fast triplet recombination  \cite{dellis2012quantum}. Furthermore, in the presence of EED interactions, the radical pair reaction is accompanied by the build-up of nuclear polarisation, which has been discussed as an enhancing factor for systems undergoing multiple photo-cycles in short succession \cite{wong2021nuclear,zarea2015spin,polenova1999coherent}. 

The evident obstacle of the EED interaction and previous suggestions of magnetoreception involving a mobile radical \cite{lee2014alternative,pooam2019magnetic,wiltschko2016light,smith2022driven} inspired us to try an elucidate if motion in the radical pair sensor can alleviate the suppressive EED effect. Specifically, we shall consider a radical pair-based model for which the inter-radical distance is time-dependent due to motion along a one-dimensional reaction coordinate. Such a model could be conceived by employing a mobile radical (in combination with the protein-bound FAD radical), if the free radical diffused in a channel or pore of cryptochrome (as illustrated in Fig.\ \ref{fig:reaction_scheme}). Alternatively, the one dimensional reaction coordinate could present itself as the effective motion of a protein-bound radical subject to protein structural rearrangements. We naively hypothesized that the resulting modulating of the inter-radical distance could potentially overcome the suppressive effect of inter-radical distances by intermittently weakening these interactions, whilst still permitting radical pair recombination in the moments of close encounters. We expected that a balance ought to be struck between the timescale and extent of radical pair motion to elicit large magnetic field effects (MFEs) whilst preventing undue spin relaxation, which is likewise induced by the modulation of the dipolar coupling \cite{kattnig2016electron}.

\section{Model}

We aim to describe magnetic field effects (MFEs) in radical pairs (RPs), the spin dynamics of which are modulated by the stochastic motion of one of the radicals. We shall assume that this motion can be described in terms of a diffusion process along a one-dimensional coordinate embedded in the protein coordinate system with a fixed direction.
The spin evolution of the RP is described by the stochastic Liouville equation for the spin density operator  $\hat{\rho}(x,t)$,
\begin{equation}
\frac{{\partial \hat \rho (x,t)}}{{\partial t}}\;\; = \;\; \left( - \text{i}\hat{\hat{L}}(x)- \hat{\hat{K}}(x) + \hat{\hat{R}} + \hat{\mathcal{L}} - k_{f} \right) \hat{\rho}(x,t),
\label{eq:SLE}
\end{equation}

\noindent where $x$ denotes the coordinate of the RP in the classical configuration space and $\hat{\mathcal{L}}$  is a functional operator describing the stochastic motion of the RP. We will find it convenient to define $x$ to represent the inter-radical distance measured from the contact distance, i.e.\ $x=0$  shall represent the radicals in contact. Describing the radical motion by a continuous diffusion processes, i.e.\ assuming the mobile radical to behave as a Brownian particle, $\hat{\mathcal{L}}$ is given by \cite{nitzan2006chemical,lukzen2006chemical}
\begin{equation}
\hat{\mathcal{L}}\hat{\rho} (x,t) = \frac{{\partial}}{{\partial x}}\left( D(x) e^{-v(x)} \frac{{\partial}}{{\partial x}} \left(e^{+v(x)}\hat{\rho} (x,t)\right)\right), 
\label{eq: BP}
\end{equation}

\noindent where $D(x)$ is the position dependent diffusion coefficient and $v(x) = u(x)/k_{B}T$  ($k_{B}$  is the Boltzmann constant, $T$  is the absolute temperature) accounts for the (possibly indirect) force interaction of the radicals expressed through potential energy $u(x)$ of the pair.

We seek solutions of eq.\ \eqref{eq:SLE} that obey the inner ($x=0$) and outer ($x=L$) boundary conditions
\begin{equation}
\hat{j}\hat{\rho}(x,t)\Bigg|_{x=0} =  D(x)e^{-v(x)}\frac{\partial}{\partial x} e^{+v(x)} \hat{\rho}(x,t)\Bigg|_{x=0} =  0
\label{eq: BC1}
\end{equation}
and
\begin{equation}
\hat{j}\hat{\rho}(x,t)\Bigg|_{x=L} =  k_{ex}\hat{\rho}(L,t),
\label{eq: BC2}
\end{equation}

\noindent where $L$ is the length of the diffusion domain and $k_{ex}$ the rate constant at which radicals reaching the outer boundary escape from the pore. In the limit $k_{ex} \xrightarrow{}\infty$, all radicals arriving at  $x=L$ escape (into the bulk), which gives way to the absorptive boundary condition $\hat{\rho}(L,t)=0$. The inner boundary, eq.\ \eqref{eq: BC1}, ensures that particles are conserved.

The spin-selective recombination of the radicals is accounted for by the superoperator $\hat{\hat{K}}(x)$. In the Haberkorn-limit of minimal singlet-triplet dephasing, $\hat{\hat{K}}(x)$ is given by \cite{haberkorn1976density,ivanov2010consistent,fay2018spin}
\begin{equation}
\hat{\hat{K}}\hat{\rho}(x,t) \;\; = \;\; \frac{1}{2} \Big\{ w_{S}(x)\hat{P}_{S} + w_{T}(x)\hat{P}_{T}, \hat{\rho}(x,t) \Big\},
\label{eq: SOP}
\end{equation}

\noindent where $\{\}$ denotes the anticommutator, and $w_{S}(x)$ and $w_{T}(x)$ are the distant dependent recombination rates for the singlet and triplet channels, respectively. $\hat{P_{S}} = \hat{\frac{1}{4}} - \hat{\mathbf{S}}_{A} \cdot \hat{\mathbf{S}}_{B}$  and  $\hat{P_{T}} = \hat{1} - \hat{P_{S}}$ are projection operators onto the singlet state and triplet manifold, respectively. The validity of this approach has been discussed widely, whereby additional reaction-induced singlet-triplet dephasing and a contribution to the exchange interaction have been highlighted \cite{haberkorn1976density,ivanov2010consistent,fay2018spin, dellis2012quantum}. Here, Haberkorn’s model is deemed appropriate as the dephasing is dominated by the hyperfine interactions and the modulation of the inter-radical interactions through radical motion, i.e.\ additional, reaction-induced dephasing can be ignored as secondary. Below, we assume that the RP recombines by bond-formation, and hence at the contact distance, exclusively in the singlet state. For this scenario the reaction rates are given by
\begin{equation}
w_{S}(x) \;= \; k_{S}\delta(x) \quad \textrm{and} \quad w_{T}(x) \;= \; 0,
\label{eq: S+T}
\end{equation}
with $\delta(x)$  denoting the Dirac delta distribution centred at the contact distance $x=0$. 

The rate constant $k_{f}$ in eq.\ \eqref{eq:SLE} accounts for the transition of the protein to a conformationally distinct state that is thought to couple the magnetic field dependent spin dynamics to downstream biological processes. In a model with immobilized radicals, which has ubiquitously been used, it is the competition between this reaction and RP recombination that accounts for the magnetoreception processes. In the scenario studied here, radical escape appears as an additional pathway.

The coherent evolution of the spin system is described by the commutator-generating superoperator, $\hat{\hat{L}}(x)$, of the RP spin-Hamiltonian, $\hat{H}(x)$, i.e.
\begin{equation}
- {\text{i}} {\hat{\hat{L}}}(x) \hat{\rho}(x,t) \;\; = \;\; -{\text{i}} \big[\hat{H}(x), \hat{\rho}(x,t)\big].
\label{eq: RP-coherent}
\end{equation}

\noindent $\hat{H}(x)$  comprises the hyperfine interactions of the radicals’ electron spins with magnetic nuclei in their surroundings, the Zeeman interaction with the geomagnetic field (assumed to have intensity $50\mskip3mu\mu$T), the exchange and the electron-electron dipolar interaction. It can be written as
\begin{equation}
{\hat{H}}(x) \;\; = \;\; \hat{H}_{A} + \hat{H}_{B} + \hat{H}_{AB}(x),
\label{eq: RP-Ham}
\end{equation}

\noindent with the free Hamiltonian of radical $i\in \{A,B\}$ being given by
\begin{equation}
{\hat{H}}_{i} \;\; = \;\; \sum_{j} \hat{\mathbf{S}}_{i} \cdot \mathbf{A}_{i,j} \cdot \hat{\mathbf{I}}_{i,j} + \mathbf{\omega} \cdot \hat{\mathbf{S}}_{i}.
\label{eq: Ham}
\end{equation}

\noindent Here, the Zeeman precession frequency is represented by $\mathbf{\omega} = -\gamma_{i}\mathbf{B}$, whereby $\gamma_{i}$  denotes the gyromagnetic ratio of the electron spin in radical $i$ and $\mathbf{B}$ the applied magnetic field, the orientation of which will be described in terms of spherical polar coordinates ($\vartheta$ and $\varphi$) in the protein coordinate system. $\mathbf{A}_{i,j}$ is the hyperfine coupling tensor between the $j$th nuclear spin and the $i$th electron spin; $\hat{\mathbf{I}}_{i,j}$ and $\hat{\mathbf{S}}_{i}$ are the corresponding vector operators of nuclear and electron spin angular momentum.
$\hat{H}_{AB}(x)$ collects the inter-radical interactions due to the exchange and electron-electron dipolar interaction, $\hat{H}_{AB}(x) = \hat{H}_{ex}(x) + \hat{H}_{eed}(x)$. The exchange interaction is accounted for by Hamiltonian
\begin{equation}
{\hat{H}}_{ex}(x) \;\; = \;\;  -J(x) \Big(\frac{\hat{1}}{2} + 2 \hat{\mathbf{S}}_{A} \cdot\hat{\mathbf{S}}_{B}\Big),
\label{eq: Hex}
\end{equation}

\noindent where $2J(x)$ is the associated spatially dependent energy difference between the singlet and triplet states. The electron-electron dipolar coupling has been considered in the point-dipole approximation,
\begin{equation}
{\hat{H}}_{eed}(x) \;\; = \;\; \frac{d_{AB}}{(x+R)^{3}} \bigg(\hat{\mathbf{S}}_{A} \cdot \hat{\mathbf{S}}_{B} - 3(\hat{\mathbf{S}}_{A} \cdot \mathbf{n})(\hat{\mathbf{S}}_{B} \cdot \mathbf{n})\bigg),
\label{eq: Heed}
\end{equation}

\noindent with the dipolar coupling constant in angular frequency units given by $d_{AB} = \mu_{0}\gamma_{A}\gamma_{B}\hbar / (4\pi)$. $\mathbf{n}$ is a unit vector parallel to the line joining the centres of the two radicals, i.e.\ parallel to the diffusion axis in the protein frame, and $R$ is the distance of the centres of spin density for the radicals in contact (i.e.\ for $x=0$). For the recombination of small molecular radicals under bond-formation, $R$ is expected to be of the order of a typical van der Waals-distance ($R=3.6\mskip3mu$\AA).

Finally, $\hat{\hat{R}}$ in eq.\ \eqref{eq:SLE} accounts for spin relaxation. For the majority of simulations, an explicit treatment of spin relaxation is omitted here for simplicity. This is a legitimate approximation provided that the RP recombines or escapes on a timescale faster than its intrinsic spin relaxation time. For some simulations, we have included uncorrelated random-field relaxation \cite{kattnig2016electron,kattnig2017radical}, as implemented by the Lindbladian term
\begin{equation}
\hat{\hat{R}}\rho (x,t) = \sum\limits_{i = 1}^2 {{\gamma _i}\left( {\sum\limits_{a \in \{ x,y,z\} } {{{\hat S}_{i,a}}\hat \rho (x,t){{\hat S}_{i,a}} - \frac{3}{4}} } \right)},
\end{equation}
with $\gamma_i$ representing the relaxation rate in radical $i$.

We here solve eq.\ \eqref{eq:SLE} subject to the initial condition
\begin{equation}
\hat{\rho}(x,0) \;\; = \;\;\frac{\hat{P}_{i}}{Tr[\hat{P}_{i}]}\delta(x),
\label{eq: init_con}
\end{equation}
where $\hat{P}_{i}\in \{\hat{P}_{S},\hat{P}_{T}\}$, depending on whether the RP is assumed to be generated from a singlet or triplet precursor. The assumption of generation at the contact distance is appropriate for formation by bond-dissociation or mildly exergonic electron transfer reactions \cite{rosspeintner2008rehm}. The recombination yield after recombination/escape of the radicals is given by 
\begin{equation}
Y_{S}(\theta, \phi) \;\; = \;\; k_{S} \int_{0}^{\infty} \operatorname{Tr}\big[\hat{P}_{S} \hat{\rho}(0,t)\big]{\text{d}}t \;\; =\;\; k_{S} Tr\big[\hat{P}_{S} \hat{\Bar{\rho}}(0)\big],
\label{eq: Rc/E_Yield}
\end{equation}
\noindent where $\hat{\overline{\rho}}(x) = \int_{0}^{\infty} \hat{\rho}(x,t)dt$ has been introduced. This quantity is most conveniently evaluated by direct time-integration of eq.\ \eqref{eq:SLE}, which yields the ordinary differential equation
\begin{equation}
\hat{\rho}(x,t=0) = \bigg[{\text{i}}\hat{\hat{L}}(x) - \hat{\hat{K}}(x) + \hat{\hat{R}} - \hat{\mathcal{L}} + k_{f}\bigg] \hat{\Bar{\rho}}(x),
\label{eq: ODE}
\end{equation}

\noindent subject to boundary conditions analogous to eqs.\ \eqref{eq: BC1} and \eqref{eq: BC2}. Eq.\ \eqref{eq: ODE} was here solved by discretizing the differential operator using finite differences and solving the resulting linear system by direct or iterative methods (biconjugate gradients method using an incomplete LU factorization as a preconditioner) \cite{werner1977theory}.  

As a measure of fidelity of the compass we use the maximal difference of the recombination yield
\begin{equation}
\Delta_{S} = {\text{max}_{\theta \phi}\;Y_{S}(\theta, \varphi)} - {\text{min}_{\theta \phi}\;Y_{S}(\theta, \varphi)}.
\label{eq: Delta}
\end{equation}

\section{Results}

We have explored the effect of one-dimensional diffusional radical motion on the magnetosensitivity in simple model systems. We commence the discussion by focusing on the popular FADH$^{\bullet}$/Z$^{\bullet}$ model, for which the roles of A$^{\bullet}$ and B$^{\bullet}$ (cf.\ Fig.\ \ref{fig:reaction_scheme}) are respectively taken on by the flavin semiquinone, FADH$^{\bullet}$, and a hypothetical radical, Z$^{\bullet}$, that is devoid of hyperfine interactions and subject to slow spin relaxation \cite{lee2014alternative}. The latter has been introduced as an idealization of the superoxide radical, O$^{\bullet-}_2$, in an attempt to sidestep the question of its fast spin relaxation, while preserving its otherwise favourable properties. As practically the hyperfine interaction structure is dominated by FADH$^{\bullet}$, the FADH$^{\bullet}$/Z$^{\bullet}$ model can serve as a good first approximation of RPs comprising various B$^{\bullet}$ with isotropic and, relative to the flavin radical, small hyperfine interactions. This condition applies to many relevant organic radicals in solution (e.g.\ the ascorbyl radical suggested in \cite{lee2014alternative} and discussed below for concreteness). 

We assume a number of properties inspired by the FADH$^{\bullet}$/O$^{\bullet-}_2$ RP. First, we assume that the recombination proceeds efficiently at the contact distance $R$ (corresponding to $x=0$) in the singlet state, in line with the process involving bond-formation (yielding the 4a-hydroperoxide) in FADH$^{\bullet}$/O$^{\bullet-}_2$. Second, we assume that the RP is formed in an overall triplet state, again at contact, thereby reflecting the picture of the O$_{2}^{\bullet-}$/FADH$^{\bullet}$ RP being formed from molecular oxygen and the fully reduced flavin co-factor FADH$^{-}$ in an electron transfer reaction that is mildly exergonic (difference of redox potentials: $\simeq$ $0.3\mskip3mu$V), and thus proceeds essentially at contact.

In line with previous findings, we find the electron-electron dipolar (EED) coupling suppresses the directional magnetosensitivity for static (i.e.\ immobilised) RPs in cryptochrome, which is again demonstrated in the Supporting Information (SI). Specifically, for a static RP with singlet-recombination of $1\mskip3mu\mu$s$^{-1}$, sizeable MFEs are only observed at large distances, as shown in Fig.\ S2 a and c. If instead the maximal electron transfer rate for any given distance is considered, effects are essentially suppressed altogether (Fig.\ S2 b and d).

To explore the overarching system behaviour, we have systematically scanned various system parameters for simple RP models for which one reaction partner is mobile and able to undergo one-dimensional diffusive excursions, as described above. The simplest model comprised only the most dominate hyperfine interaction in FADH$^{\bullet}$, namely that associated with N5 and assumed free diffusion ($v(x) = 0$). In addition, we precluded the re-entry of radicals that had escaped from the diffusion channel, which was realized by implementing an absorbing boundary condition at $L + \delta x$, where $\delta x$ is the grid spacing ($0.1\mskip3mu$\AA). Interestingly, these preliminary explorations, revealed a consistent feature that appeared essential to elicit significant directional MFEs: Sizable effects required that the RP, i.e.\ the diffusional axis, is oriented perpendicular to the flavin $\hat{\mathbf{z}}$-axis, i.e.\ within the plane of the isoalloxazine ring system of the flavin radical. All simulations supporting this conclusion are recorded in the SI (Figs.\ S4, S11-14 and S20).


\begin{figure}[tb]
	\centering
	\includegraphics[width=7.4cm]{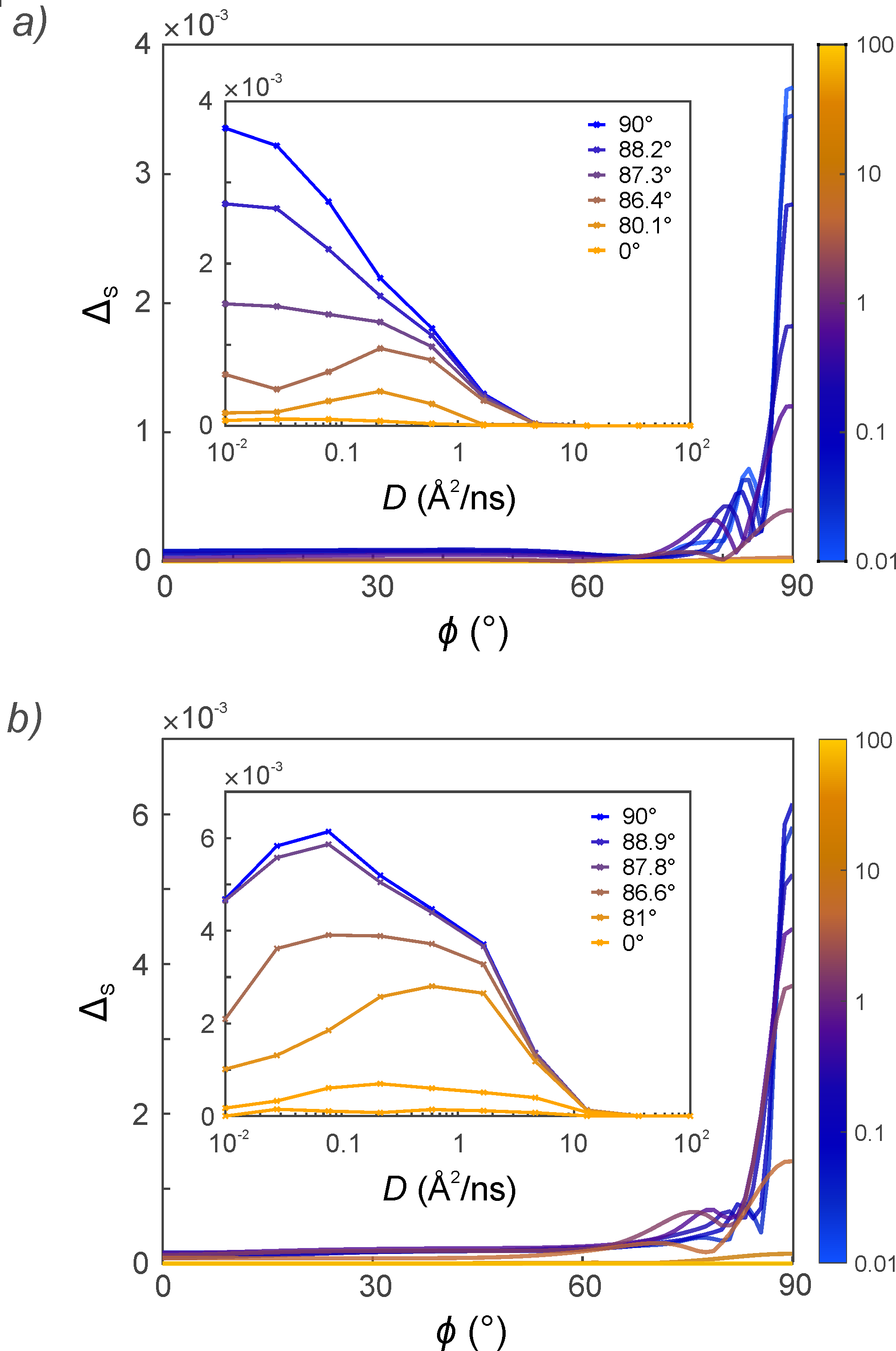}
	\caption{The spread of the recombination yield, $\Delta_{S}$, is shown  as a function of the orientation, $\phi$, of the diffusion axis in the $xz$-plane of the flavin coordinate system and, as an insert, as a function of the diffusion coefficient, $D$. The simulations assumed a) $1\mskip3mu$nm or b) $2\mskip3mu$nm long diffusion domain; radicals that approached the far end of the domain were assumed to irrevocably escape into the bulk. We have considered 1281 unique orientations for the direction of an applied magnetic field of $50\mskip3mu\mu$T flux density. Note that the second maximum at $\approx 85^{\circ}$ results merely from the definition of $\Delta_S$ and does not reflect a secondary feature. In the static limit ($D=0$), $\Delta_S=0.0204$ for all considered scenarios. Other pertinent parameters: $k_f = 0.1\mskip3mu\mu$s$^{-1}$, $J=0$.}
	\label{fig:angle}
\end{figure}

This surprising observation about the axis of diffusion holds true for the larger system including N5 and H5 for FADH$^{\bullet}$, for which too the MFE is minor for all directions deviating from colinearity with the isoalloxazine plane. Fig.\ \ref{fig:angle} illustrates this result in terms of the dependence of the directional MFE on the inclination (measured by angle $\phi$) of the diffusion axis $\mathbf{n}$ relative to the isoallozaxine plane normal ($\hat{\mathbf{z}}$) in the FADH$^\bullet$ $xz$-plane, i.e.\ $\mathbf{n} = \cos{\phi} \hat{\mathbf{z}} + \sin{\phi} \hat{\mathbf{x}}$, for various diffusion coefficients in the range $10^{-2}-100\mskip3mu$\AA$^{2}$ns$^{-1}$. Evidently, sizeable MFEs can only be elicited for $\phi \approx \pi/2$. More complex systems, specifically the system comprising N5, H5 and N10 in FADH$^{\bullet}$,  show the same result (Fig.\ S20). Based on these observations, we kept $\phi$ at a fixed value of $\pi/2$ for the remainder of the analysis.
  
Our explorations, as summarized in Fig.\ \ref{fig:angle} and the SI, further revealed that the MFE increased with decreasing diffusion coefficient. This is a consequence of assuming that the B$^\bullet$-radical irretrievably escapes into the bulk when reaching the end of the diffusion domain, $L + \delta x$.  For diffusion coefficients representative for small molecules in an aqueous solution ($\sim 100\mskip3mu$\AA$^{2}ns^{-1}$), this implies that the average residence time before escape or recombination is insufficient to yield low-field sensitivity. This is evident from the mean first-passage time of diffusive escape, which for a pore of length $L+\delta x$ is given as $T = (L + \delta x)^2/(2 D) \approx L^2/(2 D)$ (calculated from the Kramer's rate \cite{reimann1999universal, gillespie2013simple}). For $D = 100\mskip3mu$\AA$^{2}ns^{-1}$, $T\approx0.5\mskip3mu$ns, which compares unfavourably to the coherent evolution times necessary to realize magnetoreception in the geomagnetic field ($\approx 1\mskip3mu\mu$s, i.e.\ sufficient time to accommodate one Larmor precession in the applied magnetic field). On the other hand, for $D = 10^{-2}\mskip3mu$\AA$^{2}ns^{-1}$, $T\approx0.5\mskip3mu\mu$s, which predicts sizeable MFEs, which are indeed found, as shown in Fig.\ \ref{fig:angle}.

Along the same line, elongating the diffusion domain also improved the magnetosensitivity. However, as the latter is limited by the protein size, (realistically, no more than $2\mskip3mu$nm in cryptochrome or its assemblies), the realisable enhancements are expected to be small, despite the quadratic scaling of the first-passage time with $L$. Thus, free diffusion and irreversible escape appeared to strongly limit the magnetosensitivity attainable for small-molecule B$^{\bullet}$ with diffusion coefficients expected to be of the order of $10$ to $100\mskip3mu$\AA$^2$ns$^{-1}$ for cellular and aqueous environments, respectively.  In regards to this, we have considered two factors that could potentially limit radical escape and thereby boost the compass sensitivity. 

\begin{figure}
	\centering
	\includegraphics[width=6cm]{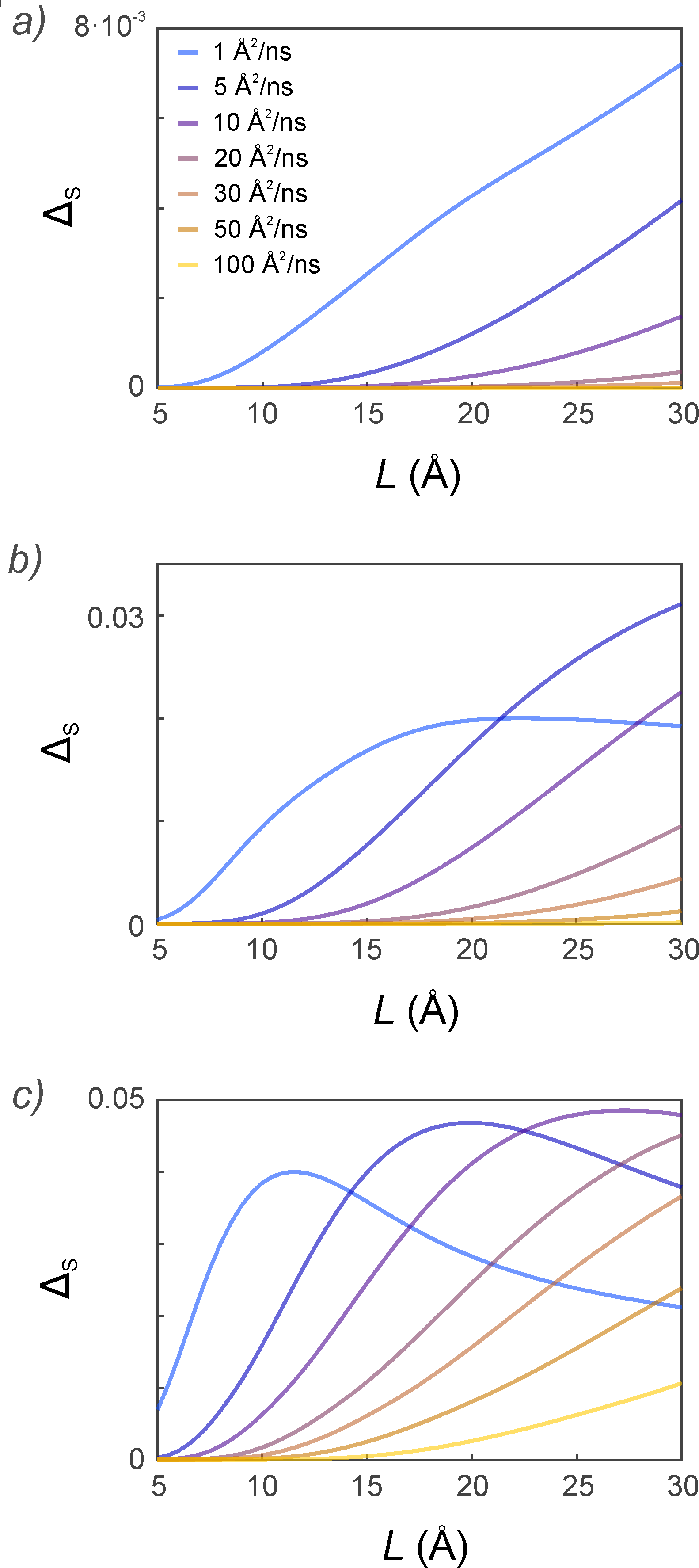}
	\caption{The anisotropy $\Delta_{S}$ is plotted as a function of the length of the one-dimensional diffusion coordinate, $L$, for three Coulomb-type interaction potentials characterized by Onsager radii, $r_{c}$, of a) $0\mskip3mu$\AA, b) $15\mskip3mu$\AA, c) $30\mskip3mu$\AA\ for various diffusion coefficients $D$ (as specified in the legend for panel a). The radical pairs diffusive excursion were assumed to proceed along the molecular $x$-axis of FAD. The model assumes radicals approaching the outer boundary escape into the bulk. The lifetime of the system has been set to $k_f^{-1} = 10\mskip3mu\mu$s. 321 unique orientations of the magnetic field have been sampled. $k_{ex}=0$, $\phi=\pi/2$, $J=0$.}
	\label{fig:potential}
\end{figure}

First, we considered binding interactions of the two radicals, i.e.\ an attractive potential, $u(x)$, that impedes the RP separation. We have assumed a Coulomb-type potential of the form $u(x)= -k_{b} T  \frac{r_{c}}{(x+R)}$, where $r_{c}$ is the Onsager radius, $r_c= \frac{e^{2}}{(4\pi\epsilon_{0} \epsilon_{r} k_{b} T)}$, here evaluating values up to $30\mskip3mu$\AA. The presence of a binding potential interaction elicits a large enhancement of the magnetosensitivity as shown in Fig.\ \ref{fig:potential}, where the directional MFE on the recombination yield $\Delta_S$ is plotted as a function of the pore length $L$ for $r_c\in \{0,15,30\}\mskip3mu$\AA.


With increasing interacting potential, the model can accommodate larger diffusion coefficients and smaller domain length, $L$, with sizeable MFEs (Fig.\ \ref{fig:potential}b, c). It is expedient to compare the MFEs so realized to the idealised, i.e.\ hypothetical, static RP model which neglected the electron-electron dipolar interaction (Fig.\ S3 in SI). For a (constant) singlet recombination rate of $k_S=1\mskip3mu\mu$s$^{-1}$, $k_f=1\mskip3mu\mu$s$^{-1}$, and otherwise comparable conditions, this system produces $\Delta_S$ = 0.0332. Remarkably, this idealized MFE is exceeded by the diffusing radical model for favourable parameters, whilst fully including the EED interactions. For example, for $r_{c}=30\mskip3mu$\AA\ and $D=5\mskip3mu$\AA$^{2}$/ns, this model produced $\Delta_{S}= 0.0399$ for $L=1.5\mskip3mu$nm, which is a 20\% increase. We also see that for $r_{c}=30\mskip3mu$\AA\ at $D=5\mskip3mu$\AA$^{2}$/ns, $\Delta_{S}= 0.0468$ for $L=2\mskip3mu$nm, which is a 41\% increase.

For the strongest bound RPs, the MFE initially increases with $L$, goes through a maximum and decreases thereafter (Fig.\ \ref{fig:potential}c). The optimal $L$ increases with increasing diffusion coefficient. These observations reflect the necessity of poising the system between RP recombination and an alternative reaction pathway, which here is escaping into the bulk, on a timescale that permits the magnetic field to impact the spin dynamics ($\approx1\mskip3mu\mu$s in the geomagnetic field), whilst being faster than spin relaxation. Again, the mean first-passage time is revealing, which in the presence of potential $u(x)=k_BTv(x)$ is given by $T= \dfrac{1}{D} \int^b_0 e^{-v(x)}\left[ \int^b_x e^{v(x)} dx' \right] dx$ with $b=L+\delta x \approx L$ \cite{nitzan2006chemical}. Indeed, the maxima of sensitivity in Fig.\ \ref{fig:potential}b and c occur for conditions for which $T\approx1\mskip3mu\mu$s (e.g.\ for $b = 20\mskip3mu$\AA\ and $r_c = 30\mskip3mu$\AA, $T$ is increased over its no-interaction limit by a factor of 30 to $T\approx1.2\mskip3mu\mu$s). Longer than necessary resident times (as resulting from further increasing $L$) are found to be detrimental, which likely results from an unbalanced recombination and escape ratio and spin relaxation, i.e.\ singlet-triplet and triplet-triplet dephasing, induced by the modulation of the inter-radical distance. The nature of these relaxation processes is analyzed in the SI in the fast-motional, Markovian limit based on Redfield theory. While this approach does not apply quantitatively here, it does reveal essential features (see SI for details). Pertaining to cryptochrome's size, we realize that small diffusion coefficients (relative to typical diffusion coefficients in water) still appear necessary to realise large magnetosensitivity when considering these parameters. Finally, the compass sensitivity also depends on the initial separation of the radical pair. As demonstrated in the SI, moderately off-contact generation of the RP only has a weak effect on the sensitivity; as the initial separation eventually approaches 
 $L$, the sensitivity expectedly drops, as the escape into the bulk again limits the lifetime (see Fig.\ S10 in the SI).

Besides binding interactions, the detrimental escape of a freely diffusing radical into the bulk could also be impeded at the transition from the diffusion domain to the bulk. In an alternative model, we have therefore assumed that the radical escape is hindered, respectively gated. To this end we have introduced a radiative outer boundary condition (for $x=L$), which is characterised by the escape rate constant $k_{ex}$. We have systematically varied $k_{ex}$, from $0.001-100\mskip3mu$\AA/ns for the freely diffusing and bound ($r_{c}>0$) RP scenarios with variable diffusivity.

\begin{figure*}[tbhp]
	\centering
	\includegraphics[width=13cm]{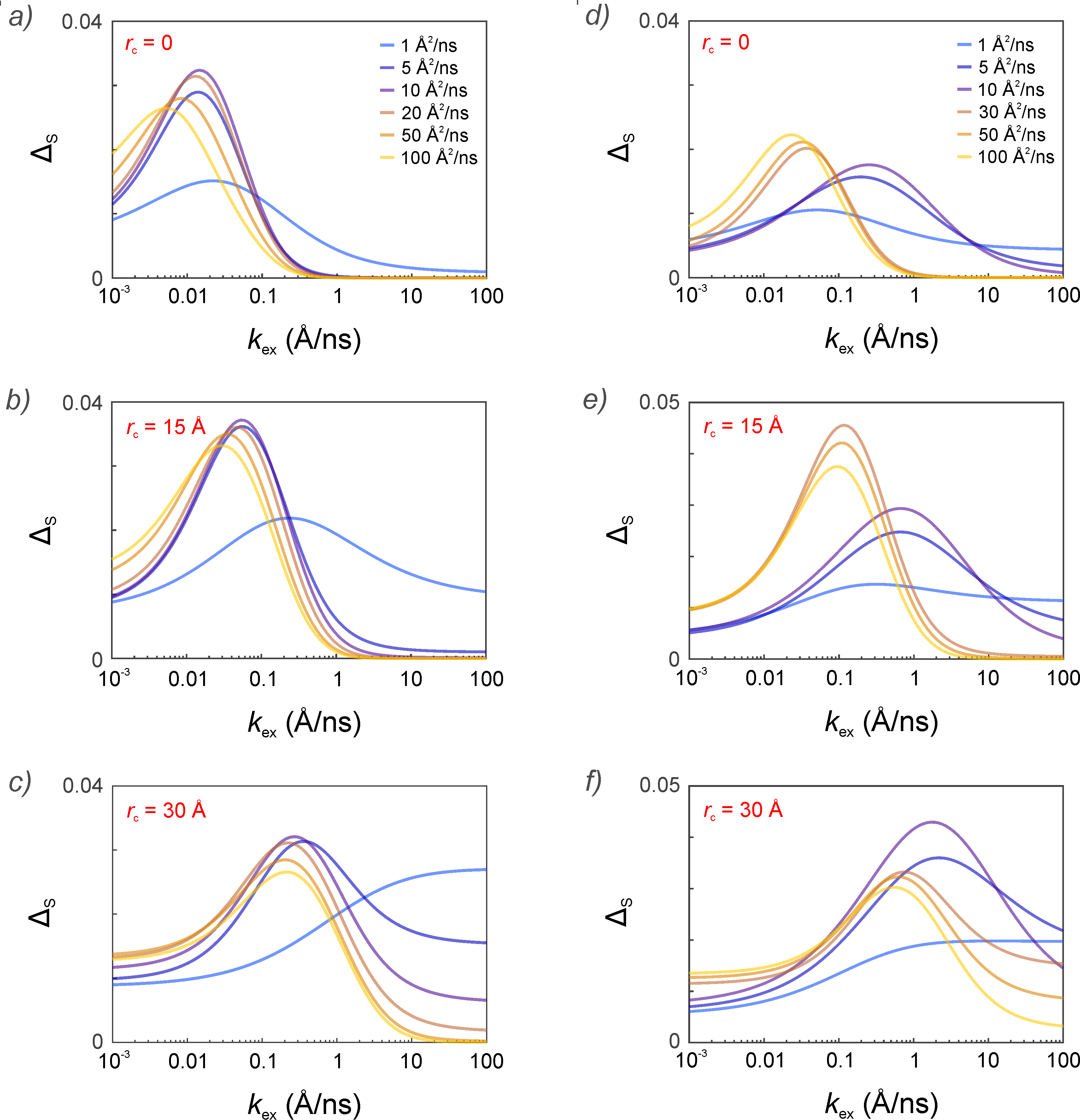}
	\caption{$\Delta_{S}$ is plotted for two fixed reaction coordinate lengths,  $L = 1\mskip3mu$nm (left) and $L=2\mskip3mu$nm (right), for three Onsager radii $r_{c}$, $0\mskip3mu$\AA, $15\mskip3mu$\AA\ and $30\mskip3mu$\AA, as a function of escape rate constant, $k_{ex}$. The exchange interaction, $J$, is neglected and $\phi$ is fixed at $\pi/2$. The directional effect has been assessed based on the three canonical magnetic field orientations, $\mathbf{B}\parallel\hat{\mathbf{x}},\hat{\mathbf{y}}, \hat{\mathbf{z}}$. $J=0$, $k_f=0.1\mskip3mu\mu$s$^{-1}$, $\phi=\pi/2$.}
	\label{fig:capping}
\end{figure*}

Fig.\ \ref{fig:capping}, left column, summarises our results for domain length of $L=1\mskip3mu$nm.  We find sizeable magnetic field sensitivity peaking for intermediate values of $k_{ex}$ for all considered scenarios.  As long as the diffusion coefficient is large enough and the diffusion domain sufficiently narrow such that the system can explore ample close and distant configurations during the time relevant to the build up of the sensitivity to the geomagnetic field (of approximately $1\mskip3mu\mu$s), the effect appeared only weakly sensitive to the diffusion coefficient considered. The MFE is then controlled by $k_{ex}$ and quickly decays as it exceeds its optimal value and the RP lifetime again becomes too short. With increasing $r_{c}$, the maximal effect is found for larger $k_{ex}$, i.e.\ the probability to reach the outer boundary is reduced by the potential interaction, thus a larger escape rate at this point can be accommodated. For slow diffusion and/or strong attraction, on the other hand, the escape into the bulk is not the limiting factor, and the MFE is sustained for large $k_{ex}$. Under the most sluggish conditions considered, large $k_{ex}$ are in fact profitable as the average radical distribution is shifted to larger distances.

To test the sensitivity of the simulations to the length of the diffusion domain (e.g.\ the pore length in cryptochrome) we also considered the case where $L=2\mskip3mu$nm. These results are summarised in the right column of Fig.\ \ref{fig:capping}. In this case, the picture is more varied whilst maintaining broad commonality in behaviour. The optimal $k_{ex}$ shifts to a larger value in correspondence to a longer domain length, which is in line with the increased dwell time to reach the outer boundary. Generally, the compass sensitivity exhibits a more pronounced tendency of levelling-off for large $k_{ex}$ for low mobility and/or strong binding. The largest effect observed was found for intermediate potential and intermediate diffusion coefficient ($D=30\mskip3mu$\AA$^{2}$/ns, $r_{c}=15\mskip3mu$\AA) producing $\Delta_{S} = 0.045$. Interestingly, the diffusion coefficient giving rise to the largest effect decreases with increasing $r_{c}$ (whilst $k_{ex}$ increases). This is likely due to increased spin relaxation at larger diffusion rates which attenuates the large effects seen for the intermediate $r_{c}$.

We have furthermore considered a larger system comprising of three hyperfine interactions, namely N5, N10, and H5 which demonstrates comparable, but unsurprisingly in view of the established literature \cite{lee2014alternative,kattnig2016electron,atkins2019optimal,gruning2022effects}, attenuated MFEs. In addition, we have studied the effect of random-field relaxation in one or both of the radicals. The model can sustain relaxation rates of up to $\gamma_i \approx 11\mskip3mu\mu$s$^{-1}$ without suffering a drop in sensitivity of more than 50 \% over its no-relaxation value if the relaxation is local to one radical. For simultaneous relaxation in both radicals, the half-sensitivity relaxation rate is given by $\gamma_1 = \gamma_2\approx 6\mskip3mu\mu$s$^{-1}$. These results are reported in the SI (Figs.\ S15 and S20).


\begin{figure}[tbhp]
	\centering
	\includegraphics[width=6cm]{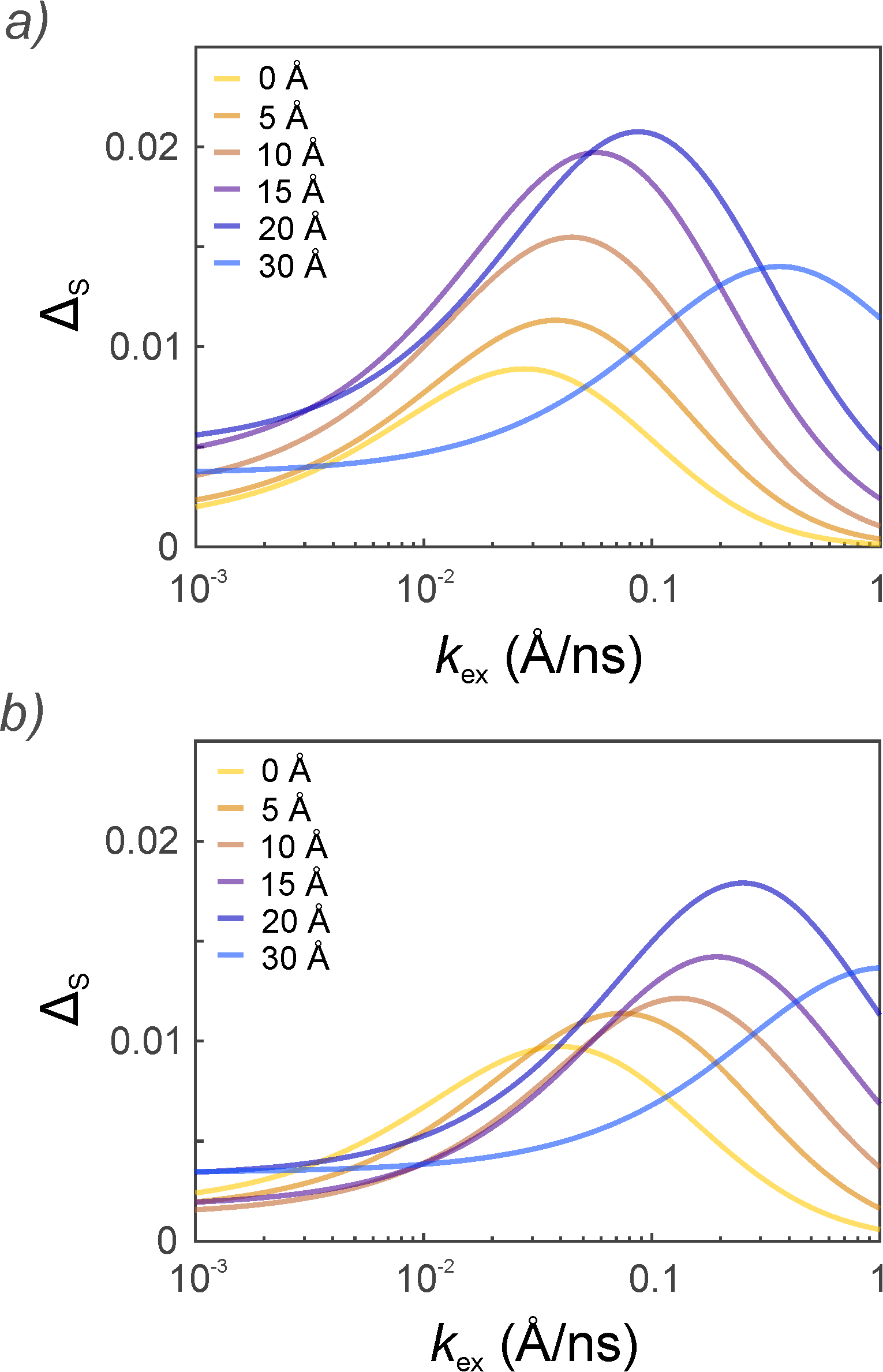}
	\caption{Directional magnetic sensitivity of a flavo semiquinone/ascorbyl radical pair. $\Delta_{S}$ is plotted as a function of the escape rate, $k_{ex}$ and Onsager radius $r_{c}$ for two separate conditions: a) $L = 1\mskip3mu$nm, $D = 10\mskip3mu$\AA$^{2}$/ns and b) $L = 2\mskip3mu$nm, $D = 30\mskip3mu$\AA$^{2}$/ns. $\phi$ is fixed at $\pi/2$, $J(x)$ is neglected. This system is composed of flavin semiquinone, modelled in terms of N5 and H5, and the ascorbyl acid radical, Asc$^{\bullet}$, modelled with H4. $k_f=0.1\mskip3mu\mu$s$^{-1}$, $\phi = \pi /2$, $\mathbf{B}\parallel\hat{\mathbf{x}},\hat{\mathbf{y}}, \hat{\mathbf{z}}$.
	}
	\label{fig:Ascorbl}
\end{figure}

Finally, in an attempt to concretize the nature of the Z$^{\bullet}$ radical, we have considered a model of the flavin/ascorbyl RP, which has been suggested in the context of the photo-reduction of cryptochrome \cite{lee2014alternative} and more recently the re-oxidation involving three radicals \cite{deviers2022anisotropic}. Based on the hyperfine structure (only one significant and small hyperfine interaction), the ascorbyl radical has been identified as a auspicious radical partner for high-sensitivity magnetoreception, which resembles the reference probe ideal \cite{lee2014alternative}. Previous model calculations, however, did not consider the EED coupling, which is only justified in the limit of fast diffusion of the radical in three dimensions. Fig.\ \ref{fig:Ascorbl} summarizes results of our model employing a one-dimensional diffusion model. The anisotropy of the singlet recombination yield is plotted as a function of the escape rate, $k_{ex}$, and the interaction strength, expressed by the Onsager radius $r_c$, for the diffusion length $L=1$ and $2\mskip3mu$nm. We achieved larger MFEs for the shorter pore length. The addition of hyperfine interactions on the partner radical, however decreases the anisotropy by 46\% for $L=1\mskip3mu$nm and 72\% for $L=2\mskip3mu$nm, relative to the same model employing the Z$^{\bullet}$ radical. The system considering a more complex expression of flavin with three hyperfine interactions, partnered with the Z$^{\bullet}$ radical, is favourable over the inclusion of the ascrobyl radical. However, they are not too far different from each other, as is evident in Fig.\ S22 in the SI. Interestingly, the dependency on $r_{c}$ is more pronounced when including a hyperfine coupling in the partner radical than in its absence. In the SI, we further study the effect of hyperfine interactions in the second radical based on the Schulten-Wolynes semiclassical model \cite{schulten1978semiclassical}. We find that the sensitivity generally decreases with increasing hyperfine-field, leveling off at a low value for effective fields exceeding $0.35\mskip3mu$mT (see Fig.\ S7 in the SI).


\section{Discussion}

\subsection{On the enablers of directional magnetosenstivity}

We have demonstrated that RP models assuming one-dimensional diffusive motion, such as would result approximately if one radical, the flavin in cryptochrome, was protein-bound and the other accessed it via a protein channel, can deliver sizable directional MFEs in the presence of the unavoidable, but oft neglected, EED interaction. As shown, large effects are subject two main requirements. First, the diffusion axes must lie within the isalloxaxine plane of the immobilized flavin radical, i.e.\ it must be perpendicular to the direction of its dominating hyperfine interaction (associated with nitrogen N5). Second, the parameters of the motion (diffusion length, diffusion coefficient, interactions) must be such that the RP lifetime is sufficient to elicit magnetosensitivity in the applied field (i.e.\ of the order of one Larmor precession or $1\mskip3mu\mu$s in the geomagnetic field). Lifetimes exceeding this value appeared non-productive, as such exacerbate the effects of spin-relaxation resulting from the modulation of the dipolar coupling by the radical motion. What enables these MFEs? It is expedient to consider a toy model that captures essential traits, while permitting analytical insights. To this end, we consider a static one-nitrogen RP, with axial hyperfine interaction characterized by $A_{xx} = A_{yy} = A_{\perp} = 0$ and non-zero $A_{zz} = A_{\parallel} \gg \omega$ ($\omega$ is the Larmor precession frequency), and EED axis parallel to the $x$-axis. The associated Hamiltonian commutes with the projection of the nuclear spin $\hat{I}_z$. This idealized picture is indeed a good approximation for the flavin N5 nucleus, the most dominant contributor to the hyperfine structure.  For a fixed inter-radical distance, and thus dipolar coupling $d$ and singlet-recombination rate $k_S$, the dynamics of this RP are dictated by the effective, non-hermitian ``Hamiltonian'' $\hat{H}_{\mathrm{eff}}(m_I)$ with blocks, labelled by the projection of the nuclear spin, $m_I$, represented by the following matrix in the basis of \{$\left| {{T_\Sigma }} \right\rangle  = \tfrac{1}{{\sqrt 2 }}\left( {\left| {{T_ + }} \right\rangle  + \left| {{T_ - }} \right\rangle } \right)$, $\left| {{T_0}} \right\rangle$, $\left| {{T_\Delta }} \right\rangle  = \tfrac{1}{{\sqrt 2 }}\left( {\left| {{T_ + }} \right\rangle  - \left| {{T_ - }} \right\rangle } \right)$, $\left| {{S}} \right\rangle$\}:
\begin{equation}
{\hat{H}_{\mathrm{eff}}(m_I)} = \frac{1}{2}\;\left( {\begin{array}{*{20}{c}}
  {-d}&0&a&0 \\ 
  0&{-d}&0&a \\ 
  a&0&{2d}&0 \\ 
  0&a&0&-i k_S 
\end{array}} \right)
\label{eq:Heff}
\end{equation}
Here, the abbreviation $a=A_{\parallel}m_I$ has been introduced and the (small) applied magnetic field and the homogeneous recombination rate $k_f$ (which affects all states equally) have been suppressed for succinctness of representation. A magnetic field applied in the $x$-direction additionally connects $\left| {{T_\Sigma}} \right\rangle$ and $\left| {{T_0}} \right\rangle$; applied along the $z$-direction, $\left| {{T_\Sigma}} \right\rangle$ and $\left| {{T_\Delta}} \right\rangle$ are coupled.

It is obvious that for a close ($d \gg A_{\parallel}$) and slowly recombining ($k_S \ll A_{\parallel},d$) RP, the singlet-triplet interconversion is impeded, i.e.\ the eigenstates comprise the approximately pure $\left| {{S}} \right\rangle$ and $\left| {{T_0}} \right\rangle$ states and no magnetic field sensitivity develops. In order to overcome this spin blockage, the radicals can obviously be farther separated, such that $d \lesssim A_{\parallel}$ and $k_S$ small, which motivates the effectiveness of (large) diffusive excursion to reinstate the magnetic field sensitivity. 
Note further that level crossings/anti-crossings could in principle give rise to specific magnetic field responses for selected distances. However, for the chosen orientation of the EED coupling no relevant crossings involving states of the same $m_I$ occur (for small $A_{\perp} \ne 0$, crossings between states of adjacent $m_I$ can be productive for $d = A_{\parallel}/\sqrt{6}$; for the EED axis parallel to the $z$-axis, relevant crossings ensue for $d=A_{\parallel}/\sqrt{3}$ and $4/3\; A_{\parallel}$ involving different and the same $m_I$, respectively). In any case, no marked level-crossing effects have become apparent for the diffusing radicals, most likely because the fleeting nature of the crossing (as is the case for RPs in free solution).

Remarkably, directional magnetic field sensitivity can be realized for this system at close contact too, provided that $k_S$ is large, i.e.\ for the RP in contact. This is a manifestation of the quantum Zeno effect, as first noted and detailedly discussed in ref.\ \citenum{dellis2012quantum}. The effect is manifest in the eigenvalues associated with the $S,T_0$-block, which, for $|m_I|=1$, are easily evaluated as 
\begin{equation}
{\lambda _{1,2}} =- \frac{1}{4}\left( {d + {\text{i}}{k_S} \pm \sqrt {4A_\parallel ^2 + {{\left( {d - {\text{i}}{k_S}} \right)}^2}} } \right).
\label{eq:Heff_lambda}
\end{equation}
In the limit of fast singlet recombination, the asymptotic dependence of the decay associated with $\lambda_1$ scales as $k_S^{-1}$, i.e.\ $\Im(\lambda_1) \sim -A_{\parallel }^2/(2 k_S)$. Thus, the decay of the associated state is slowed as $k_S$ increases, which is the characteristic feature of the Zeno effect. Dellis and Kominis argue that in the quantum Zeno regime the RP thus has enough time to ``sample'' the singlet probability, thereby paving the way to a remarkably sensitive compass. In the traditional regime with $k_S = k_T =k$, the authors found the short reaction time, $k^{-1}$, to not permit a marked increase in the singlet probability, which limited the singlet yield and compass sensitivity. However, the retarded radical recombination in the Zeno regime cannot be the (only) explanation for the results observed here, as, first, even for the homogeneous recombination case, arbitrarily small $k = k_S = k_T$ cannot reinstate the magnetosensitivity; and, second, in the Zeno scenario too the RP lifetime is upper-bounded by $k_f^{-1}$, whereby large sensitivity is realized despite the use of typical $k_f$ (i.e.\ of the same order as used for $k$ in the homogeneous case, thus precluding long-lifetime effects).

\begin{figure}[tbhp]
	\centering
	\includegraphics[width=6.9cm]{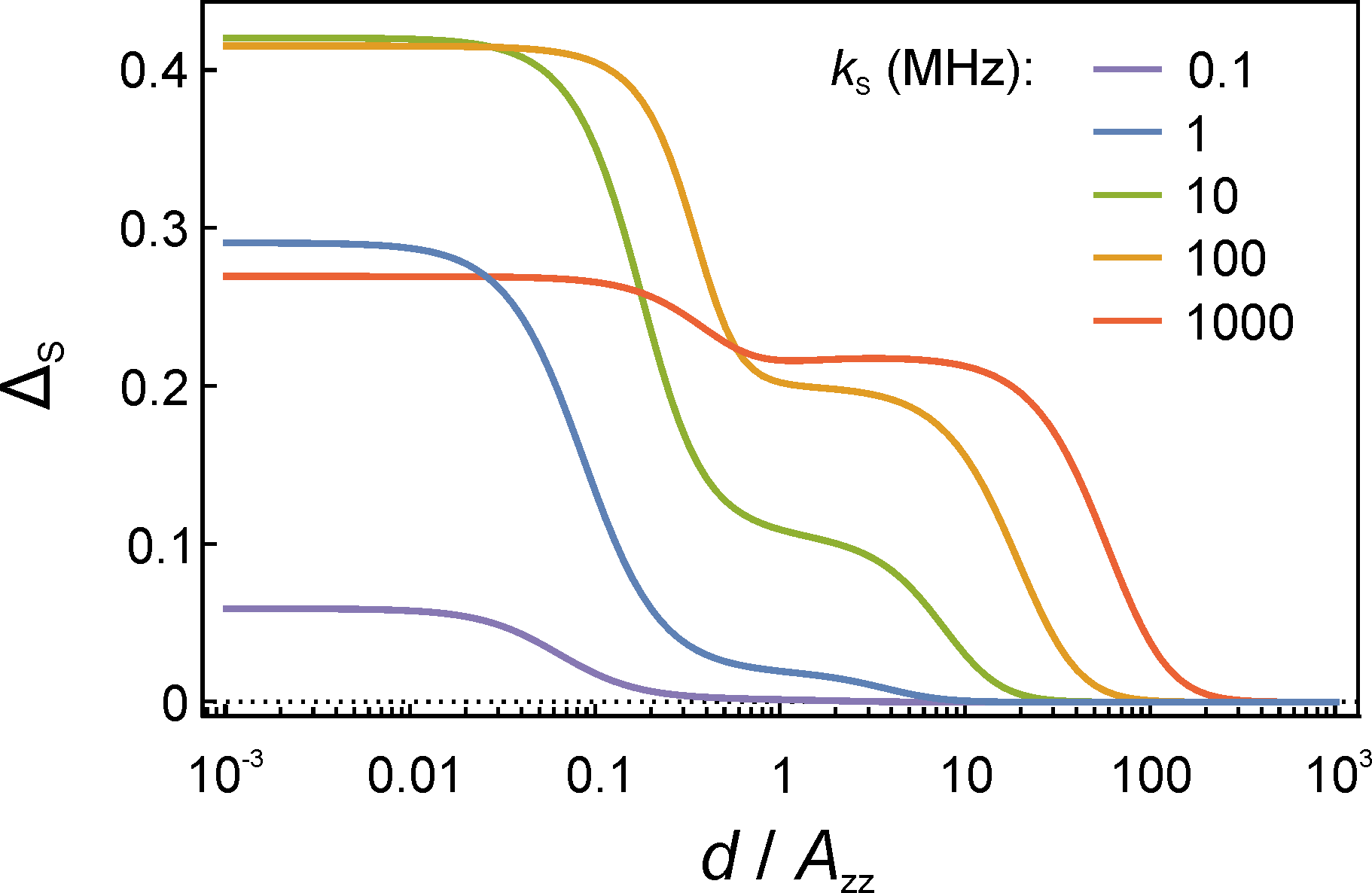}
	\caption{Directional magnetic field effect of a static toy radical pair comprising a singlet nitrogen atom coupled with axial hyperfine interaction with $A_\parallel=44\mskip3mu$MHz and $A_\perp=0$ as a function of the electron-electron dipolar coupling strength $d$ assuming that the dipolar axis corresponds to a perpendicular direction. The spin-independent decay rate constant amounted to $k_f=0.1\mskip3mu\mu$s$^{-1}$ while singlet recombination rates $k_S$ have been considered, including the quantum Zeno domain with $k_S\gg k_f$ and approximately symmetric recombination with $k_S\sim k_f$. $\Delta_S$ is the singlet recombination yield of the perpendicular minus the parallel magnetic field direction ($\omega_0/(2\pi)=1.4\mskip3mu$MHz). }
	\label{fig:zeno}
\end{figure}

Instead, we here argue that the observed large directional sensitivity in the Zeno regime results from the efficient mixing of the slowly decaying, i.e.\ Zeno-ed, state with the $T_+$,$T_-$-manifold for one field direction (perpendicular applied field; the Larmor precession connects $\left| {{T_0}} \right\rangle$ and $\left| {{T_\Sigma}} \right\rangle$) and no mixing for the other (parallel applied field; the Larmor precession connects $\left| {{T_\Sigma}} \right\rangle$ and $\left| {{T_\Delta}} \right\rangle$). As a consequence of the resulting large differential overlap of these states with the initial triplet state, the recombination yield is much larger in the former case than the latter, and a compass of large contrast becomes possible. This is enabled by the Zeno effect via the value of $\lambda_1$, but not (only) its decay (imaginary part), but importantly its associated ``energy'' (real part). Specifically, for $k_S$ large, $\Re(\lambda_1) \approx -\frac{d}{2}$, which coincidentally matches the zeroth order energies of the $\left| {{T_0 }} \right\rangle$ and $\left| {{T_\Sigma }} \right\rangle$ states (cf.\ eq.\ \eqref{eq:Heff}), thereby enabling the efficient mixing of the $T_0$,$S$-manifold and the $\left| {{T_\Sigma }} \right\rangle$ and $\left| {{T_\Delta }} \right\rangle$ states despite the small associated coupling matrix element (representing the electron spin precession in the weak geomagnetic field). For the magnetic field applied along the $z$-direction, on the other hand, $\left| {{T_\Sigma }} \right\rangle$ and $\left| {{T_\Delta }} \right\rangle$ are decoupled from $\left| {{S }} \right\rangle$ and $\left| {{T_0 }} \right\rangle$. As a result, a large contrast of the recombination yield between both orthogonal field directions and therefore compass sensitivity is realized. Fig.\ \ref{fig:zeno} illustrates the directional MFE as a function of the dipolar coupling for different values of $k_S$, including values approaching the two discussed limiting cases. We also note that, in agreement with the provided explanation, the effect is not realized with comparable compass sensitivity for singlet-born RPs, unlike the effects discussed in ref.\ \citenum{dellis2012quantum}, as the effect then loses its off-on character. In particular, for radical pairs that do not evade fast singlet recombination at contact due to low mobility, the magnetosensitivity of singlet-born radical pairs is low despite the quantum Zeno effect (see Fig.\ S9 in the SI).

For the dipolar axis parallel to $A_{\parallel}$, the effective Hamiltonian is of similar structure, but $\left| {{T_\Sigma}} \right\rangle$ and $\left| {{T_0}} \right\rangle$ are no-longer degenerate, thereby not allowing a pronounced effect of a weak magnetic field. For diffusing radicals, these delineated principles still apply, but the process is potentially complicated by the induced spin relaxation, in particular singlet-triplet dephasing \cite{kattnig2016electron}, and the possibility to induce Landau-Zener-St{\"u}ckelberg-Majorana transitions in diffusive avoided crossings \cite{smith2022driven}. Finally, it is interesting to note that for short diffusion length (e.g.\ $1\mskip3mu$nm), the observed MFEs are predominantly enabled by the quantum Zeno effect, i.e.\ for balanced recombination via the singlet and triplet channel the magnetosensitivity is strongly suppressed.

\subsection{On superoxide as radical partner}

Our model draws inspiration from the re-oxidation of cryptochrome, with a FADH$^{\bullet}$/O$_2^{\bullet-}$ RP being formed from molecular oxygen, O$_{2}$, and the fully reduced flavin co-factor. A recent re-emergence of the re-oxidation of cryptochrome in the scientific discourse has resulted from a series of \emph{in vivo} experiments suggesting dark-state, re-oxidation-related magnetosensitivity \cite{pooam2019magnetic,hammad2020cryptochrome,wiltschko2016light}.

The lack of hyperfine interactions in the superoxide would in principle provide much larger sensitivity to magnetic fields, exceeding the leading RP, flavin/tryptophan. However, it is well-known that superoxide, albeit with superior static qualities, is subject to fast spin relaxation in solution via the spin-rotational mechanism. Thus, superoxide, when allowed to tumble freely is expected to relax on a timescale proportional to the rotational correlation time, i.e.\ typically on the timescale of nanoseconds in aqueous solution, which is insufficient to establish low-field magnetosensitivity. This could be ameliorated by immobilizing the superoxide, which in view of the results developed here, is likely most easily realized directly at the reaction distance, in the flavin ring plane (but not above), with MFEs being enabled by the quantum Zeno effect, as discussed above. We remark that previous studies have suggested that the EED interaction would suppress MFEs of closely immobilized radicals \cite{efimova2008role, babcock2020electron}. However, these studies have assumed a more balanced and slower recombination in the singlet and triplet state than is necessary to ``zeno'' the system. On the other hand, if superoxide is indeed the diffusing radical, it would need to be restraint in terms of rotation mobility while retaining some form of translational mobility. One way by which such a situation could be imagined is diffusion in a narrow pore which hinders the rotation about the molecule's long axis. On the other hand, if the diffusivity is generally low, large magnetosensitivity and slow spin relaxation might naturally coincide. Specifically, the rotational correlation time $\tau_c = (6D_{\mathrm{rot}})^{-1}$ (governing spin relaxation by the spin-rotational interaction) and the translational diffusion coefficient $D$ are linked via the Stokes-Einstein and Stokes-Einstein-Debye relations, respectively. Using parameters from ref.\ \citenum{karogodina2011kinetic}, we predict that for $D=1\mskip3mu$\AA$^{2}$ns$^{-1}$, for which free diffusion gives rise to sizeable MFEs, the rotational correlation time is increased to $\tau_c\approx0.56$ns (from $2.5$ps in aqueous solution), suggesting relaxation times of hundreds of nanoseconds, thus enabling the suggested model without further assumptions.   Another possibility is that the superoxide is protein bound and its translational motion enabled by protein conformational changes rather than its own diffusion (\emph{vide supra}). It has been raised that strong binding in such a scenario will likely contribute non-negligible hyperfine couplings \cite{efimova2008role}, thereby weakening the magnetosensitivity. However, as we have shown assuming the ascobyl partner radical, this is probably tolerable for the model under question. Furthermore, here too, the binding could be weak enough to permit $\tau_c$ of the order of nanoseconds, i.e.\ fast enough to average small hyperfine interactions, yet sufficient to overcome too fast spin relaxation due to the spin rotational interaction.

The ascorbyl radical has been suggested as a substitute for Z$^{\bullet}$ that retains some of its advantages (while not being devoid of hyperfine interactions, only one isotropic hyperfine interaction coupling is significant in the freely diffusing radical), whilst relaxing slowly. While a molecular dynamics study concluded that this RP will unlikely be formed in the photoreduction of cryptochrome (as cryptochrome appears to lack dedicated ascorbic acid binding sites \cite{nielsen2017ascorbic}), it could still play a role in the reoxidation pathway, possibly in combination with the three radical model \cite{deviers2022anisotropic}. It is conceivable that the radical is in fact generated from ascorbic acid scavenging superoxide from the primary re-oxidation. In this way, the scenario of a ascorbyl radical diffusively accessing the flavin semiquinone, as considered here, could be realized. 

Finally, it is noteworthy that magnetosensitivity due to flavin/superoxide RPs have been suggested in contexts other than magnetoreception, such as hypomagnetic field effects on hippocampal neurogenesis in adult mice \cite{ramsay2022radical}, cellular bioenergetics \cite{usselman2016quantum}, and many more \cite{zadeh2022magnetic}.

\subsection{On diffusional dynamics in cryptochrome}

The environment surrounding the RP appears to play a key role in achieving sufficient MFEs. We have shown above that large magnetosensitivity requires a sufficient encounter time to manifest. Given typical diffusion rates of small molecular radicals, this typically implies that the escape into the bulk has to be impeded via an attractive interaction potential or some form of gating. Interestingly, the electrostatic potential of cryptochrome, as shown for the fully oxidized dark-state in Fig.\ \ref{fig:reaction_scheme} (the semiquinone radical form has the same charge), is markedly positive at the surroundings of a possible pore leading directly to the isoalloxazine portion of the FAD. This site could act as a potential tunnel both for oxygen diffusion and for potential radical partners, such as superoxide or the ascorbyl radical, whereby the flavin-to-tunnel orientation approximately fulfills the identified requirement of being co-planar with the isoallozaxine ring. Furthermore, the positive potential could provide the desired attraction for negatively charged radicals, such as superoxide or the ascorbyl radical.

It is furthermore worthwhile to point out that the re-oxidation of photo-reduced FAD is an integral part of the cryptochrome redox cycle (Fig.\ S1 in SI). As a consequence, access of oxygen and small molecules to the centre of the protein harboring the FAD is naturally necessary. While the re-oxidation processes has not been investigated with respect to structural details for cryptochrome \cite{pooam2019magnetic,xu2021magnetic}, growing evidence suggest that oxygen access in proteins is guided and controlled via highly specific tunnels to reach covered flavin centres \cite{baron2009multiple,saam2007molecular}. One or multiple oxygen tunnels from the outer solvent directly to the active flavin centre have now been identified for various proteins and enzymes \cite{romero2018same,piubelli2008oxygen}. Although these proteins do not closely resemble cryptochrome, they highlight common properties required for oxygen diffusion to the flavin which is essential for re-oxidation. These features included positive electrostatic potential around the flavin and hydrophobic tunnels \cite{gadda2012oxygen}. Some proteins identify a nonpolar site close to the flavin C(4a) atom, whilst other proteins relied heavily on particular amino acids \cite{piubelli2008oxygen,xu2009locally,rosini2011reaction,coulombe2001oxygen,lario2003sub}. For most cases, permanent channels are not observed in proteins, i.e.\ the transport utilizes multiple transient pathways. Future work ought to be conducted to confirm any potential tunnels in cryptochrome's dynamic ensemble, for example by implicit ligand sampling and similar methods \cite{di2015oxygen}.  Furthermore, unlike assumed in the model, diffusion channels will rarely be straight. Future studies will have to address the extent to which non-axial diffusion alters the predicted magnetosensitivity. In any case, given the near axiality of the dominant hyperfine interactions in flavins, we expect that the model will accommodate motion in the flavin ring plane with comparable sensitivity.

With respect to the possible involvement of gating at the tunnel exit to the bulk solvent, we highlight a study by Schuhmann \emph{et al.}, which has recently identified a large scale conformational fluctuation in cryptochrome 4 of pigeon that resembles a gate opening and closing to the FAD binding site in dependence on the redox state of the protein \cite{schuhmann2021exploring}. The phosphate-binding loop affects the spatial organisation of that region. Interestingly, the moving loop contains a group of arginine that could electrostatically bind to superoxide or other negatively charged radicals, thereby impeding its rotational motion, and thus spin relaxation by rotational diffusion, whilst permitting its transport in a conveyor belt-like fashion. Finally, we note that the reversible jump process of the radical, such as would result from electron transfer along a chain of redox-active residues, will result in qualitatively similar dynamics as the considered diffusion process.

\section{Conclusion}

In summary, we have considered a model employing a mobile radical approaching its radical partner, which is assumed immobilized, along a one-dimensional reaction coordinate, describing its motion by a continuous diffusion process. We have modelled a scenario relevant to cryptochrome magnetoreception with the protein-bound flavin in its semiquinone state reacting with either a hypothetical Z$^{\bullet}$ radical - an idealisation of superoxide - or the ascorbyl radical. We asked the question whether the motional dynamics of the radical can elicit sizeable directional MFEs despite the suppressive effect of the inter-radical coupling by the electron-electron dipolar interaction. Previous theoretical studies have often bypassed inter-radical interactions to search through much larger spin systems. However, the electron-electron dipolar coupling effect is an intrinsic property of radical pair system that cannot be ignored due to the necessarily close proximity of reacting radicals.

The system modelled with a diffusive radical partner was able to alleviate the suppressive effect of the EED interactions and yielded large magnetic field sensitivity for optimised parameters. Magnetosensitivity is enabled by the quantum Zeno effect, mediated through fast asymmetric recombination at the reaction distance, and/or the intermittent reduction of the EED interaction during diffusive excursions to large inter-radical distances. For short diffusion domains, the former effect is decisive. Interestingly, marked magnetosensitivity was only present when the one-dimensional reaction coordinate was oriented within the isoalloxazine-plane of the flavin co-factor, which has been rationalized in terms of the Zeeman interaction with a parallel magnetic field connecting states that are degenerate in zeroth order, thus imparting large changes in the spin dynamics. Typical diffusion rates of free, small molecules do not allow large MFEs if the diffusing radicals readily escape into the bulk at the outer boundary of the one-dimensional diffusion domain without recurrence. Large magnetosensitivity can still be enabled through an attractive interaction potential and gating at the far end of the reaction coordinate, both of which increase the radical pair lifetime by reducing the competing escape of the radical into the bulk. As discussed above, these requirements could be fulfilled for cryptochrome.

This model offers an interesting take to currently used static models of cryptochrome in magnetoreception and might be relevant in other biological radical pair reactions. The model ought to be refined in terms of realism, specifically in terms of more realistic models of the protein and approach of the diffusive radical to the FAD binding site.

\begin{acknowledgments}
We wish to acknowledge the support of the EPSRC (grant nos.\ EP/R021058/1 and EP/V047175/1), the Leverhulme Trust (RPG-2020-261) and the Office of Naval Research (ONR award number N62909-21-1-2018).
\end{acknowledgments}

\end{document}